\documentclass[sn-mathphys,Numbered]{sn-jnl}
\usepackage{multirow}%
\usepackage{amsmath,amssymb,amsfonts}%
\usepackage{amsthm}%
\usepackage{mathrsfs}%
\usepackage[title]{appendix}%
\usepackage{xcolor}%
\usepackage{textcomp}%
\usepackage{manyfoot}%
\usepackage{booktabs}%
\usepackage{algorithm}%
\usepackage{algorithmicx}%
\usepackage{algpseudocode}%
\usepackage{listings}%

\begin{document}

\title[Can black holes be a source of dark energy?]
{Can black holes be a source of dark energy?}


\author[1,2]{\fnm{Serge} \sur{Parnovsky}}\email{parnovsky@knu.ua}

\affil[1]{\orgdiv{Astronomical observatory}, \orgname{Taras Shevchenko National University of Kyiv}, \orgaddress{\street{Observatorna 3}, \city{Kyiv}, \postcode{04053}, \country{Ukraine}}}

\affil[2]{\orgdiv{D\'epartement de Physique Th\'eorique}, \orgname{Universit\'e de Gen\`eve}, \orgaddress{\street{24 quai Ernest-Ansermet}, \city{Gen\`eve}, \postcode{CH-1211}, \country{Switzerland}}}


\abstract{The hypothesis that the mass of black holes increases with time according to the same law as the volume of the part of the Universe containing it and therefore the population of BHs is similar to dark energy in its action was recently proposed. We demonstrate the reasons why it cannot be accepted, even if all the assumptions on which this hypothesis is based are considered true.}




\maketitle
\section{Introduction}
Black holes (BHs) were once considered as controversial candidates for astronomical objects. However, most astronomers have long admitted their existence. Moreover, some of them try to explain with  help of BHs two still mysterious entities, namely dark matter and dark energy (DE). The first can be considered quite reasonable, because black holes have mass and do not consist of baryons. However, at the beginning of 2023, an article \citep{p1} appeared, in which a source of DE was associated with BHs having effectively constant energy density.

In this article, we consider and reject this possibility. In doing so, we initially agree with all the assumptions proposed by \citet{p1}. Our goal is to show that we cannot explain the observed cosmological evolution even after accepting all these suppositions and hypotheses.

\section{Brief summary of hypotheses about the growth of BH masses}
At the heart of the attempt to explain DE by the influence of BH is the formula proposed in \citep{p2} and having the form
\begin{equation}
\label{e1}
M_{BH}(a)=M_0(a/a_i)^k.
\end{equation}
Here $M_{BH}$ is the mass of an individual black hole, $M_0$ is the mass of the input stellar remnant, i.e. the mass of the black hole at the time of its formation, $a$ is the current scale factor, $a_i$ is the scale factor at which the remnant was formed, and $k$ is a dimensionless constant.

Formula (\ref{e1}) raises a natural question about what is meant by the BH mass M(a). As an assumption, we assume that this quantity is the same mass that astronomers have in mind and which they estimate from astronomical observations. Therefore, it is possible to determine the value of $k$ from these observations.

For this, mass estimates for supermassive black holes (SMBHs) were used. Additionally, the masses $M_{stellar}$ of the stellar population of high-redshift and low-redshift quiescent elliptical galaxies containing these SMBHs were estimated. The values mentioned were estimated from UV / optical spectra, in particular, from luminosities and full-widths half-maximum in H$\alpha$, H$\beta$, and Mg II emission lines. The details are described in \citep{p1,p3,p4}. 

In \citep{p1}, it is stated that the offsets in stellar mass are small, and consistent with measurement bias, but the offsets in SMBH mass are much larger, reaching a factor of 7 between $z \sim 1$ and $z \sim 0$. This served as the basis for the estimates at 90\% CL
\begin{equation}
\label{e2}
k=2.96^{+1.65}_{-1.46}
\end{equation}
and
\begin{equation}
\label{e3}
k=3.11^{+1.19}_{-1.33}.
\end{equation}
These estimates are close to $k=3$ and practically exclude the case of $k=0$. 

However, there are also alternative opinions. The article \citep{p4} claims that the average BH-to-host stellar mass ratio appears to be consistent with the local value within the uncertainties, suggesting a lack of evolution of the $M_{BH}$ -- $M_{stellar}$ relation up to $z \sim 2.5$. We will not discuss the details of observations, sampling, data processing, etc. We simply note that the same data were used by \cite{p1} and \cite{p4}. Therefore, the difference in the conclusions cannot be explained by the difference in the observations or corrections used, e.g., for extinction, aperture, etc.

\section{Brief discussion of the hypothesis about the growth of BH masses}
Let us assume that the authors of \cite{p1} are right and the BH masses increase with cosmological expansion, i.e. as the scale factor $a$ increases. Let's discuss what could be causing this.

The mechanisms of BH mass increase such as accretion of surrounding matter and collapse with the formation of BHs are well known. The authors specifically consider coupling of BH. It is quite possible that there may be several SMBHs inside the galaxy that merge together.

The process of galaxies merging is well known. In this case, the mass of the stellar population of the formed galaxy can be approximately considered equal to the sum of $M_{stellar}$ of the merged galaxies. Their SMBHs coexist for a while, but may later merge. However, in all these cases, the law of conservation of energy/mass works. During the merger, the mass of the formed BH does not exceed the sum of the masses of the original BHs. The total mass of all BHs in the galaxy may decrease because of an emission of gravitational waves in the process of BH merging. The general relativity's limitation is associated only with an increase in the total area of black hole horizons.

An increase in the BH mass at accretion or collapse is compensated by a decrease in the mass of matter outside the BH. In this case,  the total mass of matter and BHs does not increase with expansion. The total mass of BHs considered separately from other types of matter (gas, dust, stars, dark matter) can increase, but it is difficult to imagine that the rate of accretion of matter on a BH is somehow related to the scale factor $a$. However, even here one can come up with a saving explanation: due to the expansion of space-time, the $a$ value increases with increasing cosmological time $t$. We can consider a monotonically increasing function $a(t)$ and an inverse one $t(a)$ and formally reduce the function $M_{BH}(t)$ to $M_{BH}(a)$.

Be that as it may, the equation (\ref{e1}) includes a scale factor $a$. Maybe the reason for the increase in mass is somehow connected with cosmology? When considering vacuum stationary solutions describing BH (Schwarzschild and Kerr metrics), space-time far from BH becomes flat and the mass of the central object can be determined from the asymptotical form of the metric. However, there are other quantities or functions associated with alternative definitions of mass. If the space-time is not asymptotically flat far from the BH, then the problem of mass determination becomes much more complicated. Authors of \cite{p1} rightly point out that we do not know a solution that describes even a Schwarzschild BH, not to mention the Kerr one, against the background of a homogeneous isotropic FLRW space-time. Let us assume that in this incomprehensible situation we can accept formula (\ref{e1}) with the value $k\simeq3$ according to (\ref{e2}) or (\ref{e3}) as a hypothesis or an empirical relationship.

But this raises a somewhat odd problem. The paper \cite{p3} compares estimates of SMBHs masses for samples with different $z$. It contains $\tau_{BH}$: the translational offset between the high- and low-redshift samples along the SMBH mass axis. According to (5) from this article $\tau_{BH}$ between the COSMOS sample (high-$z$ sample) and the low-redshift quiescent sample is equal to $1.15^{+0.25}_{-0.28}$ dex. So the SMBHs masses increase with $z$. This also follows from formula (18) from the same article \cite{p3}, according to which at 90\% confidence
\begin{equation}
\label{e4}
\frac{M_{BH}}{M_{stellar}}=(1+z)^{3.5\pm 1.4}.
\end{equation}
In this case, the mass of the stellar population $M_{stellar}$ changes much weaker than the mass of the BHs. However, this statement is directly opposite to formula (\ref{e1}). The masses of black holes were estimated from the spectrum of radiation emitted by galaxies. For the high-$z$ sample, this radiation was emitted long ago, when the scale factor of the Universe was $1+z$ times smaller. Therefore, according to (\ref{e1}), the BH masses should also be smaller.

Shortly after the article \cite{p1}, an article \citep{p5} appeared stating that the
mass functions of the two radial velocity black hole candidates in NGC 3201 place strong constraints on the cosmologically-coupled growth of black holes.

\section{DE and the BH mass growth}
Let's just discard all doubts and agree with all the assumptions of the article \cite{p1}. Let's accept formula (\ref{e1}) with $k=3$ as a hypothesis.  Will we get some analogue of TE as a result? Of course not.

The simplest version of DE -- the cosmological constant -- has a constant mass and energy density. The density of the quantity designated as $M_{BH}$ has the same property. Consider a part of space whose boundaries are fixed in the comoving coordinate system. Both $M_{BH}$ and the volume of this part increase proportionally to $a^3$. Therefore, the $M_{BH}$ density does not change with time. But this is not enough to ensure the properties of DE. It is also necessary to have a negative pressure with the equation of state (EoS) $P=-\rho$ (it corresponds to the cosmological constant) or close to it. Here $P$ is the pressure and $\rho$ is the density of mass and energy (in units with $c=1$). In the first case, the constancy of $\rho$ and $P$ is provided automatically if we proceed from the fact that DE cannot be transformed into something else and vice versa.

Within GR, both gravitational attraction and repulsion are possible. Everything is determined by the sign of the combination $\rho + 3P$. For ordinary matter, it is positive and attraction occurs. For the case of the cosmological constant or the more general case of DE, it is negative. This corresponds to gravitational repulsion or anti-gravity, which ensures the accelerated expansion of the Universe.

The black hole system does not have negative pressure. Therefore, it does not provide anti-gravity and accelerated expansion. It cannot be considered as something that works as an analogue of DE. Moreover, at present, the influence of DE prevails in the cosmological expansion, while the mass of black holes is a very small fraction of the mass of everything that fills our Universe.

But \citet{p1} came to the opposite conclusion. Here are some quotes from the article \cite{p1}: ``The redshift dependence of the mass growth implies that, at $z \lesssim 7$, black holes contribute an effectively constant cosmological energy density to Friedmann's equations. The continuity equation then requires that black holes contribute cosmologically as vacuum energy. We further show that black hole production from the cosmic star formation history gives the value of $\Omega_\Lambda$ measured by Planck while being consistent with constraints from massive compact halo objects. We thus propose that stellar remnant black holes are the astrophysical origin of dark energy, explaining the onset of accelerating expansion at $z \sim 0.7$. $<...>$  From conservation of stress-energy, this is only possible if the BHs also contribute cosmological pressure equal to the negative of their energy density, making $k \sim 3$ BHs a cosmological dark energy species. $<...>$ Taken together, we propose that stellar remnant $k = 3$ BHs are the astrophysical origin for the late-time accelerating expansion of the universe.''

There is no mention in the article \cite{p1} of the reasons why the authors came to the conclusion that the BH population has a negative pressure, and it is huge in absolute value. Indeed, without the fulfillment of condition $\rho + 3P<0$ there will be no antigravity and, accordingly, no accelerated expansion. Standard  concept of the properties of black holes rule out this possibility.

Even if we assume that our knowledge of the BH properties will change significantly in the future, they are unlikely to include negative pressure. The reason is simple. Black holes do not uniformly fill the entire space, but are concentrated into small objects. Even if they would have a negative pressure capable of providing anti-gravity, then the gravitational repulsion would be observed primarily in the region around the black hole. In this case, instead of accretion of matter from the surrounding space onto the BH, we would observe its expansion, dispersion, or flying apart, which contradicts the astronomical observations.

\section{Gravastars}
However, let's give the hypothesis another chance and assume that the BH mentioned in the article are not at all the black holes that are written about in textbooks on general relativity and whose masses are estimated from astronomical observations. The article mentions vacuum energy interior BHs, more precisely singularity-free BH models, such as those with vacuum energy interiors. In order not to confuse them with standard Schwarzschild or Kerr black holes, I will use the term gravitational vacuum condensate stars or gravastars for them. According to \citep{p6} gravastars are cold, low entropy, maximally compact objects characterized by a surface boundary layer and physical surface tension, instead of an event horizon. Within this thin boundary layer the effective vacuum energy changes rapidly, such that the interior of a non-rotating gravastar is a non-singular static patch of de Sitter space. In this case, there is negative pressure and anti-gravity, but only inside the gravastars.

Let's consider this model. Let there be only matter in the universe, which can be considered cold (its pressure is much less than density) and gravastars. Radiation makes an insignificant contribution to the total density and pressure. There is no dark energy in the model, because it is assumed that this role is played by gravastars.

Is it possible to provide a general anti-gravity, which would manifest itself in the observed accelerated expansion of the Universe? Everything depends on the sign of the combination $\rho + 3P$, which is the sum of the respective contributions of matter and gravastars. For matter it is $\rho_m$, for gravastars $\rho_{gr} + 3P_{gr}$. Since $\rho_{gr}>0$ and $P_{gr}<0$, we get $\rho_{gr} + 3P_{gr}>3P_{gr}$. For an ordinary vacuum with $P=-\rho$ this combination is equal to $2P_{gr}=-2\rho_{gr}$, but we are ready to consider the case of a non-standard vacuum inside gravastars. Anti-gravity is possible at $3P_{gr}+\rho_m<0$, that is, at $P_{gr}<-\rho_m/3$. However, even the most daring hypotheses do not suggest that black holes or gravastars provide a quarter of the total mass in the universe, and in this case we cannot get accelerated expansion.

\section{Conclusions}
Even if we accept as a hypothesis all the assumptions of paper \cite{p1}, including the formula (\ref{e1}) and the violation of conservation laws, we cannot ensure the accelerated expansion of the Universe with the help of a BH population. The latter requires an absent strong negative pressure.

{\bf Data availability statement}

No new data were created or analysed in this study.

{\bf Acknowledgement}

I am grateful to Ruth Durer both for the invitation to visit Universit\'e de Gen\`eve and work in a calm environment without air raids and blackouts, and for drawing my attention to the article \cite{p1}.

{\bf ORCID iD}

Serge Parnovsky https://orcid.org/0000-0002-1855-1404

\end{document}